\documentclass[superscriptaddress,aps,prd,showpacs]{revtex4}
\usepackage{amsfonts}
\usepackage{graphicx}
\usepackage{amsmath}
\usepackage{color}
\usepackage{soul}

\begin{document}

\setstcolor{black}

\title{Green's function approach to Chern-Simons extended electrodynamics: an effective theory describing topological insulators}
\author{A. Mart\'{i}n-Ruiz}
\email{alberto.martin@nucleares.unam.mx}
\affiliation{Instituto de Ciencias Nucleares, Universidad Nacional Aut\'{o}noma de M\'{e}xico, 04510 M\'{e}xico, Distrito Federal, M\'{e}xico}

\author{M. Cambiaso}
\affiliation{Universidad Andres Bello, Departamento de Ciencias Fisicas, Facultad de
Ciencias Exactas, Avenida Republica 220, Santiago, Chile}

\author{L. F. Urrutia}
\affiliation{Instituto de Ciencias Nucleares, Universidad Nacional Aut\'{o}noma de M\'{e}%
xico, 04510 M\'{e}xico, Distrito Federal, M\'{e}xico}

\date{\today}

\begin{abstract}
Boundary effects produced by a Chern-Simons (CS) 
extension to electrodynamics are
analyzed exploiting the Green's function (GF) method. We 
consider the
electromagnetic field coupled to a $\theta$-term in a 
way that has been
proposed to provide the correct low energy effective 
action for topological
insulators (TI). We take the $\theta$-term to be 
piecewise constant in
different regions of space separated by a common 
interface $\Sigma$, to be called the $\theta$-boundary.
Features arising due to the presence of the boundary, 
such as magnetoelectric effects, are already known in CS 
extended electrodynamics and solutions for some 
experimental setups have been found with specific 
configuration of sources. In this work we illustrate a 
method to construct the
GF that allows to solve the CS modified field equations for 
a given $%
\theta$-boundary with otherwise arbitrary configuration 
of sources. The method is illustrated by solving the
case of a planar $\theta$-boundary but can also be 
applied for cylindrical and spherical geometries for which 
the $\theta$-boundary can be characterized by a
surface where a given coordinate remains constant. The 
static fields of a point-like
charge interacting with a planar TI, as described by a 
planar discontinuity
in $\theta$, are calculated  and successfully compared 
with previously reported
results. We also compute the force between the charge 
and the $\theta$-boundary by two different methods, 
using  the energy momentum tensor approach and the 
interaction energy calculated via the GF. The infinitely 
straight current-carrying wire is also analyzed.
\end{abstract}


\maketitle

\section{Introduction}

\label{introduction}

The relevance of Chern-Simons (CS) forms \cite{Chern:1974ft} in several
branches of theoretical physics is well accounted for. In quantum field
theory in regards to anomalies \cite{ABJ} 
they played a key role and in particle physics they proved important as
well \cite{Peccei:1977hh, 'tHooft:1986nc,Wilczek:1987mv}. In general
relativity they also enjoy a prominent position as clearly reviewed
in \cite{Zanelli:2012px}. Further studies involve its uses in topological
quantum field theory \cite{Witten:1988ze}, topological string theory \cite%
{Marino:2004uf} and as a quantum gravity candidate \cite{Bonezzi:2014nua}.

In general, CS forms are amenable for capturing topological features of the
physical system  they describe, which is why in the last decade their
importance has also become apparent in the field of condensed matter physics
for describing what came to be known as topological phases. von Klitzing's
discovery of the astonishing precision with which the Hall conductance of a
sample is quantized \cite{von Klitzing:1980kg}, despite the varying
irregularities and geometry of the sample, turned out to have a topological
origin. The reason for this lies in the band structure of the sample, 
but ultimately, the Hall conductance can be expressed as an invariant
integral over the frequency in momentum space, more precisely as an integral
of the Berry curvature over the Brillouin zone \cite{SQ Shen}, inasmuch as
the genus of a manifold can be expressed in terms of an invariant integral
of the local curvature over the surface enclosing it. This quantity plays
the role of a topological order parameter uniquely determining the nature of
the quantum state, as the order parameter in Landau-Ginzburg effective field
theory determines the usual phases of quantum matter.

In this work we will be concerned with a simple case of CS theories, to
which we will refer as $\theta$-electrodynamics or simply $\theta$-ED and it
amounts to extending Maxwell electrodynamics by a parity violating term of
the form 
\begin{equation}  
\Delta \mathcal{L}_\theta = \theta (\alpha / 4 \pi^2) \mathbf{E} \cdot 
\mathbf{B} = - {\frac{\theta }{4}} (\alpha / 4 \pi^2) F_{\mu \nu} \tilde
F^{\mu \nu},\label{FF*}
\end{equation}
where $\tilde{F}^{\mu \nu }=\frac{1}{2}\epsilon ^{\mu \nu \alpha \beta }
F_{\alpha \beta }$ and $\epsilon ^{\mu \nu \alpha \beta }$ is the
Levi-Civit\`a symbol. In general $\theta$ can be a dynamical field, however,
we will take it as a constant scalar,  
making Eq.~(\ref{FF*}) a pseudo-scalar. Note that this extension is a total
derivative, producing no contribution to the field equations when usual
boundary conditions are met. If $\theta$ is not globally  constant in the manifold
where the theory is defined, then the CS-term fails to be a topological
invariant and therefore the corresponding modifications to the field
equations must be taken into consideration.

Here we will study Maxwell theory extended by Eq. (\ref{FF*}) defined on a
manifold in which 
there are two domains defined by their different constant values of $\theta$
that are separated by a common interface or boundary $\Sigma$. The $\theta$%
-term can be thought of as an effective parameter characterizing properties
of a novel electromagnetic vacuum possibly arising from a more fundamental
theory or, as applied to material media, as an effective macroscopic
parameter to describe novel quantum degrees of freedom of matter apart from
the usual permittivity $\varepsilon$ and permeability $\mu$. The former
approach has been taken in the context of classical $\theta$-ED \cite{ZH}
and in the  quantum vacuum  framework \cite{Canfora:2011fd}. For related
analyses, in several contexts, see \cite{related}. The latter approach has been used to describe
TIs. Concretely, the low-energy limit of the electrodynamics of
TIs can be described by extending Maxwell electrodynamics by Eq. (\ref{FF*}%
), originally formulated in 4+1 dimensions but appropriately adapted to
lower dimensions by dimensional reduction \cite{Qi:2008ew}. Thus, $\theta$%
-ED as a topological field theory (TFT), serves as model for many
theoretical \cite{theo_topoins} and experimental realizations for studying
detailed properties of topological states of quantum matter %
\cite{tech-apps, topo_reviews}. 

The formulation of $\theta $-ED pursued in this work can be considered as a particularly simple version of  the so called Janus field theories \cite{CFKS,DEG,Chen, Witten2, Kim1, Kim2}. Generally speaking, such theories are characterized by having space-time dependent coupling constants, as is $\theta$ in our model. They have been actively explored in the context of the AdS/CFT correspondence. Nevertheless, as we have  already mentioned,  in the case of  $\theta $-ED this idea is applied to a simpler but more realistic system, that constitutes an effective low energy theory which allows to compute the response of topological insulators to arbitrary external sources and currents in a planar geometry, with direct extensions  to cylindrical and spherical geometries. Janus field  theories were motivated, from the gravitational sector of the AdS/CFT correspondence,  by an exact and  non-singular  solution for the dilatonic field in  type II-B supergravity,  which was found  in a simple deformation of the $AdS_{5}\times S^{5}$ geometry \cite{Bak}. Even though the solution breaks all the original supersymmetries it proves to be stable under a large class of perturbations  \cite{Bak,FNSS,CCDAVPZ}. The dilaton acquires a constant value at the boundary, where $AdS_{5}$ is recovered, but adopts different values at each half of the boundary. On the other hand, the AdS/CFT correspondence requires the existence of a dual gauge theory on the boundary for every non-singular solution of type IIB supergravity with appropriate asymptotics which in this case is a four dimensional $\mathcal{N}=4$ SYM theory living in the boundary \cite{Bak}. In other words, a running dilaton induces space-time dependent coupling constants in the gauge theories in the dual sector, which defines what is called a Janus field theory. In our case, the four dimensional $\mathcal{N}=4$ SYM theory is replaced by the CS modified ED, where we take the electromagnetic coupling to be globally  constant, while the topological coupling to the Pontryagin invariant has different  constant values  at each side of a plane interface and  suffers a jump across such boundary. In relation to $\theta $-ED, it is interesting to recall that the authors of Ref. \cite{CFKS} proposed a model for the dual theory arising from  the Janus solution, where the $\mathcal{N}=4$ SYM coupling $g(z)$ affects only the kinetic term of the non-Abelian gauge field, together with the interaction terms in the original Lagrangian for the standard $\mathcal{N}=4$ SYM  theory. The model completely breaks the $16$ original supersymmetries of the $\mathcal{N}=4$ SYM theory. Moreover,  $g(z)$ is taken as constant on each side of a planar interface $(z=0)$, with a sharp jump across it. In this way, the gauge field part of the action is the non-abelian generalization of the Maxwell action in an inhomogeneous medium with permitivity $\epsilon $ and permeability $\mu $ related by $\epsilon (z)=1/\mu (z)=1/g^{2}(z)$. The YM fields satisfy boundary conditions at the interface, which are derived by integrating the equations of motion over the standard infinitesimal pill-shaped regions across the boundary, in a way similar to standard electrodynamics. The YM Green's function is also obtained  by  using image methods. Nevertheless,  let us emphasize  that this  model does not include a coupling to the YM Pontryagin invariant, so its Abelian limit does not reproduce $\theta $-ED. The inclusion of the topological coupling $\theta (z)$ in addition to the YM coupling $g(z)$ is developed in Refs. \cite{Witten2, Kim2},  where 1/2 BPS vacuum configurations are studied in particular. As shown in Ref. \cite{Witten2} half of the original supersymmetries can be maintained provided such couplings are constrained by the relations $1/g^{2}(z)=D\sin 2\psi (z)$ and $\theta (z)=\theta_{0}+8\pi ^{2}D\cos 2\psi (z)$.  The case of a sharp interface respecting the above  constraints is  also considered in Ref. \cite{Kim2} and it is studied in the abelian Coulomb phase, by setting two different constant values $\psi _{1}$ and $\psi _{2}$ at each side of a planar boundary. However, in the case of $\theta $-ED, supersymmetry does not enter and we are choosing the electromagnetic coupling to be globally constant, \textit{i.e.}  $g_{1}=g_{2}=e$, while only the topological coupling $\theta (z)$ becomes discontinuous at the sharp boundary, with constant values $\theta_{1}\neq \theta _{2}$ in each side. As can be seen already, these two systems are not equivalent and we will later  discuss this issue in more detail.

The paper is organized as follows. In section \ref{model} we review the
basics of Chern-Simons electrodynamics defined on a four dimensional
spacetime in which the $\theta $-value is piecewise constant in different
regions of space separated by a common boundary $\Sigma $. 
 In section \ref{Green}  we restrict ourselves to the static case and we construct
the  GF matrix for the planar geometry corresponding to
a $\theta $-boundary located at $z=a$. Section \ref{applications} is devoted
to different applications, \emph{e.g.}, the problems of a point-like charge
and a current-carrying wire near a
planar $\theta $-boundary. The interaction energy (and forces) between a
charge-current distribution and a $\theta $-interface is briefly discussed.
Contact between the results obtained with our method and others in the
existing literature is made.  A
concluding summary of our results comprises the last section \ref{summary}.
Throughout the paper, Lorentz-Heaviside units are assumed ($\hbar =c=1$),
the metric signature will be taken as $\left( +,-,-,-\right) $ and the
convention $\epsilon ^{0123}=+1$ is adopted.

\section{$\protect\theta$-Electrodynamics in a bounded region}

\label{model}

Our model is based on Maxwell electrodynamics coupled to a gauge invariant $\theta$-term as described by the following action
\begin{equation}
\mathcal{S}=\int_{\mathcal{M}}d^{4}x\left[ -\frac{1}{16\pi }F_{\mu \nu
}F^{\mu \nu }-\frac{1}{4}{\theta }\frac{\alpha }{4\pi^2 }F_{\mu
\nu }\tilde{F}^{\mu \nu }-j^{\mu }A_{\mu }\right] ,  \label{action}
\end{equation}%
where $\alpha =e^{2}/\hbar c$ is the fine structure constant and $j^{\mu }\;$is a conserved
external current.
The coupling constant for the $\theta$-term, $\alpha / 4 \pi ^{2}$, is
chosen in such a way that the total electric charge $q 
_{e} = \frac{1}{4 \pi} \int d\textbf{S} \cdot \textbf{D}$ 
has to be an integer multiple of the electron charge $e$, 
whereas the magnetic charge $q _{m} = \frac{1}{4 \pi} 
\int d\textbf{S} \cdot \textbf{B}$ should be an integer multiple of $g 
= e / 2 \alpha$ by the Dirac 
quantization condition \cite{Wilczek:1987mv}. 
Later we will recall the reasoning which shows that, quantum mechanically, the allowed values of $\theta$ 
are $0$ or $\pi$ (mod $2 \pi$).

\begin{figure}[tbp]
\begin{center}
\includegraphics
{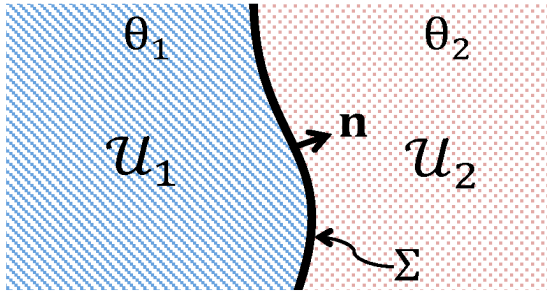}
\end{center}
\caption{{\protect\small Region over which the electromagnetic field theory is
defined.}}
\label{regions}
\end{figure}

The $(3+1)$-dimensional spacetime is $\mathcal{M}=\mathcal{U}\times \mathbb{R%
}$, where $\mathcal{U}$ is a three-dimensional manifold and $\mathbb{R}$
corresponds to the temporal axis. We make a partition of space in two
regions: $\mathcal{U}_{1}$ and $\mathcal{U}_{2}$ in such a way that
manifolds $\mathcal{U}_{1}$ and $\mathcal{U}_{2}$ intersect along a common
two-dimensional boundary $\Sigma $, to be called the $\theta $-boundary, so
that $\mathcal{U}=\mathcal{U}_{1}\cup \mathcal{U}_{2}$ and $\Sigma =\mathcal{%
U}_{1}\cap \mathcal{U}_{2}$, as shown in Fig. \ref{regions}. We also assume
that the field $\theta $ is piecewise constant in such way that it takes the
constant value $\theta =\theta _{1}$ in region $\mathcal{U}_{1}$ and the
constant value $\theta =$ $\theta _{2}$ in region $\mathcal{U}_{2}$. This
situation is expressed in the characteristic function 
\begin{equation}
\theta \left( \textbf{x} \right) =\left\{ 
\begin{array}{c}
\theta _{1}\;\;\;,\;\;\; \textbf{x} \in \mathcal{U}_{1} \\ 
\theta _{2}\;\;\;,\;\;\; \textbf{x} \in \mathcal{U}_{2}%
\end{array}%
\right. .  \label{theta}
\end{equation}%
In this scenario the $\theta $-term in the action fails to be a global
topological invariant because it is defined over a region with the boundary $%
\Sigma $. Varying the action gives rise to a set of Maxwell equations with
an effective additional current density with support at the boundary 
\begin{equation}
\partial _{\mu }F^{\mu \nu }=\tilde{\theta}\delta \left( \Sigma \right)
n_{\mu }\tilde{F}^{\mu \nu }+4\pi j^{\nu },  \label{FieldEqs}
\end{equation}%
where $n_{\mu }$ is the outward normal to $\Sigma $, and $\tilde{\theta}%
= \alpha \left( \theta _{1}-\theta _{2}\right) /\pi $, which enforces the invariance of the classical dynamics under the shifts of $\theta$ by any constant, $\theta \rightarrow \theta + C$. 
Current conservation
can be verified directly by taking the divergence at both sides of Eq.~(\ref%
{FieldEqs}) and  using symmetry properties.
The set of Eqs.~(\ref{FieldEqs}) for $\theta $-ED can be written  as 
\begin{eqnarray}
\nabla \cdot \mathbf{E} &=&\tilde{\theta}\delta \left( \Sigma \right) 
\mathbf{B}\cdot \mathbf{n}+4\pi \rho ,  \label{GaussE} \\
\nabla \times \mathbf{B}-\frac{\partial \mathbf{E}}{\partial t} &=&\tilde{%
\theta}\delta \left( \Sigma \right) \mathbf{E}\times \mathbf{n}+4\pi \mathbf{%
J},  \label{Ampere}
\end{eqnarray}% 
while the homogeneous equations are included in the Bianchi identity $\partial _{\mu }\tilde{F}^{\mu \nu }=0$.
Here $\mathbf{n}$ is the unit normal to $\Sigma $ shown in the Fig.~\ref%
{regions}. In this work we consider a simple  geometry corresponding to
a  surface $\Sigma$ taken as the plane $z=a$.

As we see from Eqs. (\ref{GaussE}-\ref{Ampere}) the behavior of $\theta $-ED
in the bulk regions $\mathcal{U}_{1}$ and $\mathcal{U}_{2}$ is the same as
in standard electrodynamics. The $\theta $-term modifies Maxwell equations
only at the surface $\Sigma $. Here $F^{i0}=E^{i}$, $F^{ij}=-\varepsilon
^{ijk}B^{k}$ and $\tilde{F}^{i0}=B^{i}$, $\tilde{F}^{ij}=\varepsilon
^{ijk}E^{k}$. Equations (\ref{GaussE}-\ref{Ampere}) also suggest that the
electromagnetic response of a system in the presence of a $\theta $-term can
be described in terms of Maxwell equations in matter 
\begin{equation}
\nabla \cdot \textbf{D} = 4 \pi \rho \;\;\;\ , \;\;\;\ \nabla \times \textbf{H} = 4 \pi \textbf{J} , \label{MaxEqMatter}
\end{equation}
with constitutive
relations 
\begin{equation}
\mathbf{D}=\mathbf{E}+\frac{\alpha }{\pi }\theta \left(z\right) \mathbf{B}%
,\;\;\;\;\;\;\;\;\;\mathbf{H}=\mathbf{B}-\frac{\alpha }{\pi }\theta \left(
z\right) \mathbf{E},  \label{CONST_REL}
\end{equation}%
where $\theta \left(z\right) $ is given in Eq. (\ref{theta}). If $\theta (z)$ is globally constant in $\mathcal{M}$, there is no
contribution to Maxwell equations from the $\theta $-term in the action,
even though $\theta$ still is present in the constitutive relations.
In fact, the additional contributions of a globally constant $\theta$ to each of the modified Maxwell equations,  (\ref{GaussE}) and (\ref{Ampere}) cancel due to the homogeneous equations.

Now we return to the problem of the allowed values of $
\theta$ to describe topological insulators. $U(1)$ gauge 
theories with nonzero $\theta$ ($\theta$-ED) exhibit an $SL(2, 
\mathbb{Z})$ duality group which strongly constrains the 
quantum physics \cite{Witten3,Karch}. This group is obtained by repeated 
applications of the $S$ and $T$ generators of electric-magnetic duality. 
The $S$ generator is associated with 
the invariance of classical Maxwell equations in matter (\ref{MaxEqMatter}) 
(supplemented with magnetic charge and current 
densities) under duality rotations. Only the special case of 
a duality transformation by $\pi / 2$ is consistent with 
the requirement that the electric charge and the 
magnetic charge are quantized.

The aforementioned rescaling symmetry $\theta \to \theta +C$ would allow to set $\theta$ to zero at the classical level. Quantum mechanically, however, given that for properly quantized electric and magnetic fluxes $S _{\theta} / \hbar$ is an integer multiple of $\theta$, only $C = 2 \pi n$ for integer $n$ is an allowed symmetry, otherwise non-trivial contributions to the path integral would result. Furthermore, since $\mathbf{E}\cdot \mathbf{B}$ is odd under $t \to -t$, only $\theta = 0$ and $\theta = \pi$ give a time-reversal symmetric theory. Thus, time reversal takes $\theta$ into $- \theta$, so $\theta = 0$ is time reversal invariant per se, whereas $\theta = \pi$ is invariant after the shift $\theta \rightarrow \theta + 2 \pi$. This is typically referred as the $T$ generator of the electric-magnetic duality. The two transformations $S$ and $T$ generate the $SL(2, \mathbb{Z})$ symmetry group acting  on the fields \cite{Karch}.

Next we study the effects of a $\theta$-interface in the propagation of the fields. Assuming that the time derivatives of the fields are finite in the vicinity
of the surface $\Sigma $, the field equations imply that the normal component
of $\mathbf{E}$, and the tangential components of $\mathbf{B}$, acquire
discontinuities additional to those produced by superficial free charges and
currents, while the normal component of $\mathbf{B}$, and the tangential
components of $\mathbf{E}$, are continuous. For  vanishing free
sources on the surface the boundary conditions read
\begin{eqnarray}
 \mathbf{E}_{z}\big|^{z=a^+}_{z=a^-} &=&\tilde{\theta}\mathbf{B}_{z}\big|%
_{z=a},\;\;\;\;\;\;\;\;\; \mathbf{B}_{\parallel }\big|^{z=a^+}_{z=a^-}=-\tilde{%
\theta}\mathbf{E}_{\parallel }\big|_{z=a},  \label{Ampere-BC} \\
 \mathbf{B}_{z}\big|^{z=a^+}_{z=a^-} &=&0,\;\;\;\;\;\;\;\;\;\;\;\;\;\;\;\; \;\;\;\; \mathbf{E}%
_{\parallel }\big|^{z=a^+}_{z=a^-}=0.  \label{Faraday-BC}
\end{eqnarray}
The continuity conditions  (\ref{Faraday-BC}) imply
that the right hand sides of Eqs. (\ref{Ampere-BC}) are 
well defined
and they represent self-induced surface charge and 
current densities, respectively. An
immediate consequence of the boundary conditions 
(\ref{Ampere-BC}-\ref{Faraday-BC}) is 
that the presence of a
magnetic field crossing the surface $\Sigma $ is sufficient 
to generate an
electric field, even in the absence of free electric charges.  
Many interesting magnetoelectric effects due 
to a 
$\theta$-boundary have 
been highlighted using different approaches. For 
example, 
electric charges close to 
the interface $\Sigma$ induce magnetic mirror 
monopoles (and vice versa) \cite{science, Kim1, Kim2}. Also, the 
propagation of electromagnetic waves across a 
$\theta$-boundary have been studied finding that a non 
trivial 
Faraday rotation of the polarizations appears \cite{ZH, Kim1, Kim2, Hehl}. It is worth mentioning that 
with the modified boundary conditions, several properties 
of conductors still hold for
static fields as far as the
conductor does not lie in the $\Sigma $-boundary, in 
particular, conductors
are equipotential surfaces and the electric field just 
outside the conductor
is normal to its surface.

\section{The Green's function method}

\label{Green} 

In this section we use the GF method to solve
static boundary-value problems in $\theta $-ED in terms of the electromagnetic
potential $A^{\mu}$. Certainly one could solve for the electric and
magnetic fields from the modified Maxwell equations together with the
boundary conditions (\ref{Ampere-BC}-\ref{Faraday-BC}), however,
just as in ordinary electrodynamics, there might be occasions where
information about the sources is unknown and rather we are provided with
information of the 4-potential at the given boundaries. In these cases, the
GF method provides the general solution, to a given boundary-value problem (Dirichlet or Neumann) for arbitrary sources. Nevertheless, in the sequel 
we  restrict ourselves to contributions of free sources outside the $\theta$-boundary, and without boundary conditions imposed on additional surfaces, except for the standard boundary conditions at infinity.  

Since the homogeneous Maxwell equations that express the relationship between potentials and fields are not modified, the electrostatic and magnetostatic fields
can be written in terms of the $4$-potential $A ^{\mu} = \left( \phi , 
\mathbf{A} \right)$ according to $\mathbf{E} = - \nabla \phi $ and $\mathbf{B%
} = \nabla \times \mathbf{A}$ as usual. In the Coulomb gauge $\nabla \cdot 
\mathbf{A} = 0 $, the $4$-potential satisfies the equation
of motion 
\begin{equation}
\left[ - \eta ^{\mu} _{\; \nu} \nabla ^{2} - \tilde{\theta}\delta \left(
z-a\right) \epsilon _{\;\;\;\;\ \nu }^{3 \mu \alpha }\partial _{\alpha }%
\right] A^{\nu }=4 \pi j^{\mu },  \label{FiedlEqPlaneConfig}
\end{equation}
together with the boundary conditions 
\begin{equation}
A ^{\mu} \big| _{z= a^{-}} ^{z = a^{+}} = 0 \;\;\;\;\ , \;\;\;\;\ \left(
\partial _{z} A ^{\mu} \right) \big| _{z= a^{-}} ^{z = a^{+}} = - \tilde{\theta}
\epsilon ^{3 \mu \alpha} _{\;\;\;\;\ \nu} \partial _{\alpha} A ^{\nu } \big| %
_{z=a} .  \label{BC4Pot}
\end{equation}
One can further check that these boundary conditions for the $4$-potential 
corresponds to the ones
obtained in Eqs. (\ref{Ampere-BC}-\ref{Faraday-BC}). 

To obtain a general solution for the potentials $\phi$ and $\mathbf{A}$ in the presence of arbitrary external sources $j^\mu (\textbf{x})$, we
introduce the GF $G ^{\nu} _{\; \sigma} \left( \mathbf{x} , 
\mathbf{x} ^{\prime} \right)$ solving Eq. (\ref{FiedlEqPlaneConfig}) for a
point-like source, 
\begin{equation}
\left[ - \eta ^{\mu} _{\; \nu} \nabla ^{2} - \tilde{\theta}\delta \left(
z-a\right) \epsilon _{\;\;\;\;\ \nu }^{3 \mu \alpha }\partial _{\alpha }%
\right] G ^{\nu} _{\; \sigma} \left( \mathbf{x} , \mathbf{x} ^{\prime}
\right) = 4 \pi \eta ^{\mu} _{\; \sigma} \delta ^{3} \left( \mathbf{x} - 
\mathbf{x} ^{\prime} \right) ,  \label{EqsGreenMatrix}
\end{equation}
together with the boundary conditions (\ref{BC4Pot}), in such a way that the
general solution for the $4$-potential in the Coulomb gauge is 
\begin{equation}
A ^{\mu} \left( \mathbf{x} \right) = \int d ^{3} \mathbf{x} ^{\prime} \; G
^{\mu} _{\; \nu} \left( \mathbf{x} , \mathbf{x} ^{\prime} \right) j^{\nu}
\left( \mathbf{x} ^{\prime} \right) .  \label{GreenMatrix}
\end{equation}
According to Eqs. (\ref{EqsGreenMatrix}) the diagonal entries of the GF matrix are related with the electric and magnetic fields arising from the charge and current density sources, respectively, although they acquire $\theta$-dependence. However, the non-diagonal terms encode the magnetoelectric effect, \textit{i.e.} the charge (current) density contributing to the magnetic (electric) field.

In the following we concentrate in constructing the solution to Eq. (\ref{EqsGreenMatrix}). The GF we consider has
translational invariance in the directions parallel to $\Sigma $, that is in
the transverse $x$ and $y$ directions, while this invariance is broken in the $z$ direction. Exploiting this symmetry we further introduce the Fourier
transform in the direction parallel to the plane $\Sigma$, taking the
coordinate dependence to be $\left( \mathbf{x}-\mathbf{x}^{\prime }\right)
_{\parallel }$ $=(x-x^{\prime },\;y-y^{\prime })$ and define 
\begin{equation}
G _{\;\nu }^{\mu }\left( \mathbf{x} , \mathbf{x} ^{\prime }\right) = 4\pi
\int \frac{d^{2}\mathbf{p}}{\left( 2\pi \right) ^{2}}e^{i\mathbf{p}\cdot
\left( \mathbf{x}-\mathbf{x}^{\prime }\right) _{\parallel }} g_{\;\nu }^{\mu
}\left( z,z^{\prime }\right) ,  \label{RedGreenDef}
\end{equation}%
where $\mathbf{p}=(p_{x},p_{y})$ is the momentum parallel to the plane $%
\Sigma $ \cite{CED}. In Eq. (\ref{RedGreenDef}) we have suppressed the
dependence of the reduced GF $g_{\;\nu }^{\mu }$ on 
 $\mathbf{p}$.

Due to the antisymmetry of the Levi-Civit\`a symbol, the partial derivative
appearing in the second term of the GF Eq. (\ref%
{EqsGreenMatrix}) does not introduce  derivatives with respect to $z$, but only
in the transverse coordinates. This allows us to write the full reduced
GF equation as 
\begin{equation}
\left[ \partial ^{2}\eta _{\;\nu }^{\mu }+i\tilde{\theta}\delta \left(
z-a\right) \epsilon _{\;\;\;\;\ \nu }^{3 \mu \alpha }p_{\alpha }\right]
g_{\;\ \sigma }^{\nu }\left( z,z^{\prime }\right) =\eta _{\;\ \sigma }^{\mu
}\delta \left( z-z^{\prime }\right) ,  \label{RedGreenFunc}
\end{equation}%
where $\partial ^{2}= \mathbf{p} ^{2}-\partial _{z}^{2}$, $p^{\alpha
}=\left( 0 , \mathbf{p} \right) $ and $\mathbf{p} ^{2} = - p ^{\alpha} p
_{\alpha}$. The solution to Eq. (\ref{RedGreenFunc}) is simple, but not straightforward. For solving it we employ a method similar to that used for
obtaining the GF for the one-dimensional $\delta$-function
potential in quantum mechanics, where the free GF is used for
integrating the GF equation with $\delta$-interaction. To this end we consider a reduced free
GF having the form $\mathcal{G}_{\;\nu }^{\mu }\left(
z,z^{\prime }\right) =\mathfrak{g}\left( z,z^{\prime }\right) \eta _{\;\nu
}^{\mu }$, associated with the operator $\partial ^{2}$ previously defined,
that solves 
\begin{equation}
\partial ^{2}\mathcal{G}_{\;\nu }^{\mu }\left( z,z^{\prime }\right) =\eta
_{\;\nu }^{\mu }\delta \left( z-z^{\prime }\right), \label{RedGreenFuncA1}
\end{equation}%
satisfying the standard boundary conditions at infinity, where
\begin{equation}
\mathfrak{g}(z,z^{\prime })=\frac{1}{2p}e^{-p|z-z^{\prime }|} \label{RFSGF}
\end{equation}
and $p = \vert \textbf{p} \vert$. Note that Eq. (%
\ref{RedGreenFuncA1}) demands the derivative of $\mathfrak{g}$ to be
discontinuous at $z=z^{\prime}$, \emph{i.e.,} $\partial _{z} \mathfrak{g} \left( z ,
z ^{\prime} \right) \big|  _{z= z^{\prime -}} ^{z=z^{\prime +}} = -1$, and then
the continuity of $\mathfrak{g}$ at $z=z^{\prime}$ follows \cite{CED}.

Now we observe that Eq. (\ref{RedGreenFunc}) can be   directly  integrated by
using the free GF Eq. (\ref{RedGreenFuncA1}) together with the
properties of the Dirac delta-function, thus reducing the problem to a set
of coupled algebraic equations, 
\begin{equation}
g_{\;\ \sigma }^{\mu }\left( z,z^{\prime }\right) =\eta _{\;\sigma }^{\mu }%
\mathfrak{g}\left( z,z^{\prime }\right) -i\tilde{\theta}\epsilon _{\;\;\;\;\
\nu }^{3 \mu \alpha }p_{\alpha }\mathfrak{g}\left( z,a\right) g_{\;\ \sigma
}^{\nu }\left( a,z^{\prime }\right) .  \label{RedGreenFuncA3}
\end{equation}
Note that the continuity of $\mathfrak{g}$   at $z=z'$  implies the continuity of $g_{\;\
\sigma }^{\mu }$, but the discontinuity of $\partial _{z} \mathfrak{g}$ at the same point 
 yields 
\begin{equation}
\partial _{z} g_{\;\ \sigma }^{\mu } \left( z , z^{\prime }\right) \big| %
_{z= a ^{-}} ^{z = a^{+}} = -i \tilde{\theta}\epsilon _{\;\;\;\;\ \nu }^{3 \mu
\alpha }p_{\alpha } \partial _{z} \mathfrak{g}\left( z,a\right) \big| _{z=
a^{-}} ^{z = a^{+}} g_{\;\ \sigma }^{\nu }\left( a,z^{\prime }\right) = i \tilde{%
\theta}\epsilon _{\;\;\;\;\ \nu }^{3 \mu \alpha } p_{\alpha } g_{\;\ \sigma
}^{\nu }\left( a,z^{\prime }\right) ,  \label{BCRedGreenFunc}
\end{equation} 
from which the boundary conditions for the 4-potential in Eq. (\ref{BC4Pot}) 
are recovered. In this way the solution (\ref{RedGreenFuncA3}) guarantees that the
boundary conditions at the $\theta$-interface are satisfied.

Now we have to solve for the various components $g_{\;\ \sigma }^{\mu }$. To
this end we split Eq. (\ref{RedGreenFuncA3}) into $\mu = 0$ and $\mu =
j=1,2,3 $  components; 
\begin{eqnarray}
g ^{0} _{\; \sigma} \left( z , z ^{\prime} \right) &=& \eta ^{0} _{\;
\sigma} \mathfrak{g} \left( z , z ^{\prime} \right) - i \tilde{\theta}
\epsilon ^{3 0 i} _{\;\;\;\ j} p _{i} \mathfrak{g} \left( z , a \right) g
^{j} _{\; \sigma} \left( a , z ^{\prime} \right) ,  \label{g0S} \\
g ^{j} _{\; \sigma} \left( z , z ^{\prime} \right) &=& \eta ^{j} _{\;
\sigma} \mathfrak{g} \left( z , z ^{\prime} \right) - i \tilde{\theta}
\epsilon ^{3 j i} _{\;\;\;\ 0} p _{i} \mathfrak{g} \left( z , a \right) g
^{0} _{\; \sigma} \left( a , z ^{\prime} \right) .  \label{gjS}
\end{eqnarray}
 Now we set $z = a$ in Eq. (\ref{gjS}) and then substitute
into Eq. (\ref{g0S}) yielding 
\begin{equation}
g ^{0} _{\; \sigma} \left( z , z ^{\prime} \right) = \eta ^{0} _{\; \sigma} 
\mathfrak{g} \left( z , z ^{\prime} \right) - i \tilde{\theta} \epsilon ^{3
0 i} _{\;\;\;\ j} p _{i} \eta ^{j} _{\; \sigma} \mathfrak{g} \left( z , a
\right) \mathfrak{g} \left( a , z ^{\prime} \right) - \tilde{\theta} ^{2} 
p ^{2} \mathfrak{g} \left( z , a \right) \mathfrak{g} \left( a , a
\right) g ^{0} _{\; \sigma} \left( a , z ^{\prime} \right) ,  \label{g0S-2}
\end{equation}
where we  use the result $\epsilon ^{3 0 i} _{\;\;\;\ j} \epsilon ^{3 j
k} _{\;\;\;\ 0} p _{k} p _{i} = p ^{2}$. Solving for $g ^{0} _{\; \sigma} \left( a , z ^{\prime}
\right)$ by setting $z=a$ in Eq. (\ref{g0S-2}) and inserting the result back in that equation, we obtain 
\begin{equation}
g ^{0} _{\; \sigma} \left( z , z ^{\prime} \right) = \eta ^{0} _{\; \sigma} 
\left[ \mathfrak{g} \left( z , z ^{\prime} \right) + \tilde{\theta} p ^{2} \mathfrak{g} \left( a , a \right) A \left( z , z ^{\prime} \right) %
\right] + i \epsilon ^{30i} _{\;\;\;\ \sigma} p _{i} A \left( z , z
^{\prime} \right) ,  \label{g0S-3}
\end{equation}
where
\begin{equation}
A\left( z,z^{\prime }\right) =-\tilde{\theta}\frac{\mathfrak{g}\left(
z,a\right) \mathfrak{g}\left( a,z^{\prime }\right) }{1+ p ^{2} 
\tilde{\theta}^{2}\mathfrak{g}^{2}\left( a,a\right) }.  \label{A(Z,Z)}
\end{equation}
The remaining components can be obtained by substituting $g ^{0} _{\;
\sigma} \left( a , z ^{\prime} \right)$ in Eq. (\ref{gjS}). The result is 
\begin{equation}
g ^{j} _{\; \sigma} \left( z , z ^{\prime} \right) = \eta ^{j} _{\; \sigma} 
\mathfrak{g} \left( z , z ^{\prime} \right) + i \epsilon ^{3 j k} _{\;\;\;\
0} p _{k} \left[\eta ^{0} _{\; \sigma} - i \tilde{\theta} \epsilon ^{3 0 i}
_{\;\;\;\ \sigma} p _{i} \mathfrak{g} \left( a , a \right) \right] A \left(
z , z ^{\prime} \right) .  \label{gjS-2}
\end{equation}
Equations (\ref{g0S-3}) and (\ref{gjS-2}) allow to write the general
solution as 
\begin{equation}
g_{\;\nu }^{\mu }\left( z,z^{\prime }\right) =\eta _{\;\ \nu }^{\mu }%
\mathfrak{g}\left( z,z^{\prime }\right) +A \left( z,z^{\prime }\right)
\left\lbrace \tilde{\theta}\mathfrak{g}\left( a,a\right) \left[ p^{\mu}
p_{\nu } + \left( \eta _{\;\ \nu }^{\mu } + n^{\mu }n_{\nu } \right) p ^{2} \right] +i \epsilon _{\;\ \nu }^{\mu \;\ \alpha 3}p_{\alpha}
\right\rbrace ,  \label{GenSolGreenPlaneConf}
\end{equation}%
where $n _{\mu} =\left( 0,0,0,1\right) $ is the normal to $\Sigma $%.

The reciprocity between the position of the unit charge and the position at
which the GF is evaluated $G_{\mu \nu }(\mathbf{x},\mathbf{x}
^{\prime }) = G_{\nu \mu }(\mathbf{x} ^{\prime},\mathbf{x})$ is one of its most remarkable properties. 
From Eq. (\ref%
{RedGreenDef}) this condition demands
\begin{equation}
g_{\mu \nu }(z,z^{\prime }, \mathbf{p})=g_{\nu \mu }(z^{\prime },z,- \mathbf{%
p}),
\end{equation}%
which we verify directly from Eq. (\ref{GenSolGreenPlaneConf}).
The symmetry $g_{\mu \nu }\left( z,z^{\prime }\right) =g_{\nu \mu
}^{\ast }\left( z,z^{\prime }\right) =g_{\mu \nu }^{\dagger }\left(
z,z^{\prime }\right) $ is also manifest.

The various components
of the static GF matrix in coordinate representation are obtained by computing the Fourier
transform defined in Eq. (\ref{RedGreenDef}), with the reduced GF given by Eq. (\ref{RFSGF}). The details are presented in Appendix \ref{IntegralsStaticCase}. The
final results are
\begin{eqnarray}
G_{\;0}^{0}\left( \mathbf{x},\mathbf{x}^{\prime }\right) &=&\frac{1}{|%
\mathbf{x}-\mathbf{x}^{\prime }|}-\frac{\tilde{\theta}^{2}}{4+\tilde{\theta}%
^{2}}\frac{1}{\sqrt{R^{2}+Z^{2}}},  \label{G00} \\G_{\; i}^{0}\left( \mathbf{x},\mathbf{x}^{\prime }\right) &=& - \frac{2 \tilde{\theta}}{4+\tilde{\theta}^{2}} \frac{\epsilon _{0ij3} R ^{j}}{R ^{2}}\left( 1-\frac{Z}{\sqrt{R^{2}+Z^{2}}}\right) ,  \label{G0i} \\ G _{\; j}^{ i }\left( \mathbf{x},\mathbf{x}^{\prime }\right) &=& \eta ^{i} _{\; j} G _{\; 0}^{ 0}\left( \mathbf{x},\mathbf{x}^{\prime }\right) - \frac{i}{2}\frac{\tilde{\theta}^{2}}{4+ \tilde{\theta}^{2}} \partial ^{i} K _{j} \left( \mathbf{x},\mathbf{x}^{\prime }\right) , \label{Gij}
\end{eqnarray}
where $Z=| z - a | + | z^{\prime } - a |$, $R ^{j} = \left( \mathbf{x-x}^{\prime }
\right) _{\parallel } ^{j}  = \left( x - x^{\prime } , y - y^{\prime} \right)$,
$R=|\left(  \mathbf{x-x}^{\prime }\right) _{\parallel }|\;$and%
\begin{equation}
K ^{j} \left( \mathbf{x},\mathbf{x}^{\prime }\right) =2i\frac{\sqrt{%
R^{2}+Z^{2}}-Z}{R^{2}} R ^{j} .
\end{equation}
Finally, we observe that Eqs. (\ref{G00}-\ref{Gij}) contain all the required
elements of the GF matrix, according to the choices of $z$ and $z^{\prime
}$ in the function $Z$.

\section{Applications}

\label{applications}

\subsection{Point-like charge near a planar $\theta$-boundary}

\label{charge_near_plane}

Let us consider a point-like electric charge $q$ located at a distance $b>0$ from
the $z=0$ plane, where we have chosen $a=0$. Also, the region $z<0$ is
filled with a topologically non-trivial insulator  whereas the region $%
z>0$ is  the vacuum ($\theta_2=0$). For simplicity we choose
the coordinates such that $x^{\prime }=y^{\prime }=0$. Therefore, 
the
current density is $j^\mu(\mathbf{x}{}^{\prime })=q \eta ^{\mu} _{\; 0} \delta \left( x^{\prime
}\right) \delta \left( y^{\prime }\right) \delta \left( z^{\prime }-b\right) 
$. According to Eq. (\ref%
{GreenMatrix}), the solution for this problem is 
\begin{equation}
A^{\mu }\left( \mathbf{x}\right) =qG_{\;0}^{\mu }\left( \mathbf{x},\mathbf{r}%
\right) ,  \label{SolPointCharge}
\end{equation}%
where $\mathbf{r}=b\hat{\mathbf{e}}_{z}$. We first study the
electrostatic potential. From Eq. (\ref{G00}),
\begin{eqnarray}
z>0\;\;\;\;\ &:&\;\;\;\;\ G_{\;0}^{0}\left( \mathbf{x},\mathbf{r}\right) =%
\frac{1}{|\mathbf{x}-\mathbf{r}|}-\frac{\tilde{\theta}^{2}}{4+\tilde{\theta}%
^{2}}\frac{1}{|\mathbf{x}+\mathbf{r}|},  \label{G00z>} \\
z<0\;\;\;\;\ &:&\;\;\;\;\ G_{\;0}^{0}\left( \mathbf{x},\mathbf{r}\right) =%
\frac{4}{4+\tilde{\theta}^{2}}\frac{1}{|\mathbf{x} - \mathbf{r}|} .
\label{G00z<}
\end{eqnarray}%
For $z>0$ the GF   yields  the electric potential  $%
A^{0}\left( \mathbf{x}\right) =qG_{\;0}^{0}\left( \mathbf{x},\mathbf{r}%
\right)$ which can be interpreted as due to two point-like electric charges, one of strength $q$ at $\mathbf{%
r}$, and the other, the image charge, of strength $-q\tilde{\theta}^{2}/(4+%
\tilde{\theta}^{2})$, at the point $-\mathbf{r}$. For $z<0$ only one point-like
electric charge appears, of strength $4q/(4+\tilde{\theta}^{2})$ located at $\mathbf{r}$.

>From Eq. (\ref{SolPointCharge}) we see that two components of the magnetic
vector potential are nonzero, $A^{1}\left( \mathbf{x}\right)
=qG_{\;0}^{1}\left( \mathbf{x},\mathbf{r}\right) $ and $A^{2}\left( \mathbf{x%
}\right) =qG_{\;0}^{2}\left( \mathbf{x},\mathbf{r}\right) $. The
corresponding GF  components for each region are given by 
\begin{eqnarray}
G_{\;0}^{1}\left( \mathbf{x},\mathbf{r}\right) &=&-\frac{2\tilde{\theta}}{4+%
\tilde{\theta}^{2}}\frac{y}{R^{2}}\left\{ 
\begin{array}{c}
1-\frac{z+b}{|\mathbf{x}+\mathbf{r}|}\;\;\;\ ,\;\;\;\ z>0 \\ 
1+\frac{z-b}{|\mathbf{x}+\mathbf{r}|}\;\;\;\ ,\;\;\;\ z<0%
\end{array}%
\right.
\label{G01} \\
G_{\;0}^{2}\left( \mathbf{x},\mathbf{r}\right) &=&+\frac{2\tilde{\theta}}{4+%
\tilde{\theta}^{2}}\frac{x}{R^{2}}\left\{ 
\begin{array}{c}
1-\frac{z+b}{|\mathbf{x}+\mathbf{r}|}\;\;\;\ ,\;\;\;\ z>0 \\ 
1+\frac{z-b}{|\mathbf{x}+\mathbf{r}|}\;\;\;\ ,\;\;\;\ z<0%
\end{array}%
\right. \label{G02}
\end{eqnarray}%
according to Eqs. (\ref{G0i}). It is difficult to
interpret the components of the vector potential directly, however the
magnetic field $\mathbf{B}=\nabla \times \mathbf{A}$ is illuminating. In fact
\begin{eqnarray}
z>0\;\;\;\;\ &:&\;\;\;\;\ \mathbf{B}\left( \mathbf{x}\right) =\frac{2q\tilde{%
\theta}}{4+\tilde{\theta}^{2}}\frac{\mathbf{x}+\mathbf{r}}{|\mathbf{x}+%
\mathbf{r}|^{3}}, \\
z<0\;\;\;\;\ &:&\;\;\;\;\ \mathbf{B}\left( \mathbf{x}\right) =\frac{-2q%
\tilde{\theta}}{4+\tilde{\theta}^{2}}\frac{\mathbf{x}-\mathbf{r}}{|\mathbf{x}%
-\mathbf{r}|^{3}}.
\end{eqnarray}%
Thus we observe that the magnetic field for $z>0$ is that due to a magnetic monopole
of strength $2q \tilde{\theta} / ( 4 + \tilde{\theta}^{2}) $ located at $- 
\mathbf{r}$. For $z<0$ we have a magnetic monopole of strength -$2q\tilde{%
\theta} / ( 4 + \tilde{\theta}^{2})$ located at $\mathbf{r}$.

The solution shows that, for an electric charge near the planar surface of a
topological insulator, both an image magnetic charge and a image electric
charge will be induced. The appearance of magnetic monopoles in this
solution seems to violate the Maxwell law $\nabla \cdot \mathbf{B}=0$,
which remained unaltered in the case of $\theta$-ED. Nevertheless, recalling that $\left( \mathbf{x}\pm \mathbf{r}%
\right) /|\mathbf{x}\pm \mathbf{r}|^{3}\sim \mathbf{\nabla }_{x}(1/|\mathbf{x%
}\pm \mathbf{r}|)$, we have $\nabla \cdot \mathbf{B\sim \nabla }_{x}^{2}(1/|%
\mathbf{x}\pm \mathbf{r}|) \sim \delta (\mathbf{x} \pm \mathbf{r})$ in a
region where $\mathbf{x\neq }\pm \mathbf{r}$. Physically, the magnetic field
is induced by a surface current density 
\begin{equation}
\mathbf{J}=\tilde{\theta}\delta \left( z\right) \mathbf{E}\times \mathbf{n}=%
\frac{4q\tilde{\theta}}{4+\tilde{\theta}^{2}}\frac{R}{\left(
R^{2}+b^{2}\right) ^{3/2}} \delta \left( z\right) \hat{\varphi},
\end{equation}
that is circulating around the origin. However such induced field has the
correct magnetic field dependence expected from a magnetic monopole. This
current is nothing but the Hall current \cite{science}.

It is worth mentioning that these results were also obtained using different approaches. On the one hand the authors in Ref. \cite{science} used the image method to show that an electric charge near a topological surface state induces an image magnetic monopole due to the magneto-electric effect and, of course, emphasized the possible experimental verification via a gas of quantum particles carrying fractional statistics, consisting of the bound states of the electric charge and the image monopole charge. 

At this stage we clarify  the differences between the $\theta$-ED approach we  are following and the $1/2$ BPS construction in the sharp interface discussed in Ref. \cite{Kim2}. As we mentioned in the Introduction the 8 remaining supersymmetries in the latter case are enforced by demanding the couplings to be related in the following way
\begin{equation}
\frac{1}{e^2}= D \sin 2\psi(z), \qquad \theta=\theta_0+ 8 \pi^2 D \cos 2\psi,
\label{SUSYCONST}
\end{equation}  
where one chooses the constant values  $\psi_1$ and $\psi_2$  for $z>0$ and $z<0$, respectively. The constraint (\ref{SUSYCONST}) does not allow to simultaneously set $e_1=e_2$ and $\theta_1\neq \theta_2$, which corresponds to the case of $\theta$-ED, where supersymmetry is irrelevant. In other words, the limit $g=0$ in  the electric and magnetic fields  of the single dyon  at $z=a$ (Eqs. (5.10) of Ref. \cite{Kim2}), which were  calculated using the method of images, do not reproduce the corresponding fields obtained from our Eqs. (\ref{G00z>}), (\ref{G01}) and (\ref{G02}). Also, the transmitted an reflected fields of  massless waves propagating across the interface reported in Ref. \cite{Kim2} do not correspond to those calculated for $\theta$-ED in Refs. \cite{Hehl,ZH}. It is worth recalling that these couplings enter through the complexified paramenter $\tau=\theta/2\pi+ 4\pi i/g^2$, which is familiar in the study of the action of the group $SL(2, \mathbb{Z})$ on a topological insulator with nontrivial permitivity, permeability and $\theta$-angle \cite{Karch}.

\subsection{Force between a charge and a planar $\theta$-boundary}

\label{force}

In this section we formulate the interaction energy and the forces arising
between external sources and a TI as represented by $\theta$-boundary with a planar
symmetry. We use both, the GF matrix and the stress-energy tensor.

The interaction energy between a charge-current distribution and a
topological insulator is 
\begin{equation}
E_{int}=\frac{1}{2}\int d\mathbf{x}\int d\mathbf{x}^{\prime }j^{\mu }\left( 
\mathbf{x}\right) \left[ G_{\mu \nu }\left( \mathbf{x},\mathbf{x}^{\prime
}\right) -\eta _{\mu \nu }\mathcal{G}\left( \mathbf{x},\mathbf{x}%
^{\prime }\right) \right] j^{\nu }\left( \mathbf{x}^{\prime }\right) ,
\end{equation}%
where $\mathcal{G}\left( \mathbf{x},\mathbf{x}^{\prime }\right) = 1 /
\vert \mathbf{x} - \mathbf{x} ^{\prime} \vert $ is the GF in
vacuum. The first contribution represents the total energy of a
charge-current distribution in the presence of the $\theta$-boundary,
including mutual interactions. We evaluate this energy for the case
considered in the previous subsection of a point-like electric charge at position 
$\mathbf{r}=b \hat{\mathbf{e}} _{z}$. Making use of Eq. (\ref{G00z>}), the interaction energy is
\begin{equation}
E_{int}=\frac{q^{2}}{2}\left[ G_{00}\left( \mathbf{r},\mathbf{r}\right) -%
\mathcal{G} \left( \mathbf{r},\mathbf{r}\right) \right] =-\frac{q^{2}}{2%
}\frac{\tilde{\theta}^{2}}{4+\tilde{\theta}^{2}}\frac{1}{2b} .
\end{equation}
Our result implies that the force on the charge exerted by the $\theta$-boundary is
\begin{equation}
\mathbf{F} = - \frac{\partial E_{int}}{\partial b} \hat{\mathbf{e}} _{z}=-%
\frac{q^{2}}{\left( 2b\right) ^{2}} \frac{\tilde{\theta}^{2}}{4+\tilde{\theta%
}^{2}} \hat{\mathbf{e}} _{z} ,  \label{force1}
\end{equation}
noting that it is always attractive. This can be interpreted as the force between the charge $q$ and the image
charge $-q\tilde{\theta}^{2}/(4+\tilde{\theta}^{2})$ according to Coulomb law.

The field point of view provides an alternative derivation of this result.
To compute the force on the charge we calculate the normal component of the
flow of momentum into the $\theta $-boundary. In terms of the stress-energy
tensor this force is 
\begin{equation}
\mathbf{F}=-\hat{\mathbf{e}}_{z}\int_{\Sigma ^{+}}dST_{zz}\left( \Sigma
^{+}\right) ,  \label{INTST}
\end{equation}%
where the integration is over the surface $\Sigma ^{+}$, just outside the $%
\theta $-interface, at $z=0^{+}$. The identification of the stress tensor in
the case of $\theta $-electrodynamics proceeds along the standard lines of
electrodynamics in a medium (see for example Ref.\cite{CED}), where we read the
rate at which the electric field does work on the free charges 
\begin{equation}
\mathbf{J\cdot E}=-\mathbf{\nabla \cdot }\left( \frac{1}{4\pi }\mathbf{%
E\times H}\right) -\frac{1}{4 \pi}\left( \mathbf{E\cdot }\frac{\partial \mathbf{D%
}}{\partial t}+\mathbf{H\cdot }\frac{\partial \mathbf{B}}{\partial t}\right)
\label{ECONS}
\end{equation}%
and the rate at which momentum is transferred to the charges%
\begin{equation}
\left( \rho \mathbf{E}+\mathbf{J\times B}\right) _{k}=-\frac{%
\partial }{\partial t}\left( \frac{1}{4\pi }\mathbf{D\times B}\right) _{k}-%
\frac{1}{4\pi }\left[ D_{i}\partial _{k}E_{i}-\partial _{i}\left(
D_{i}E_{k}\right) \right] -\frac{1}{4\pi }\left[ B_{i}\partial
_{k}H_{i}-\partial _{i}\left( B_{i}H_{k}\right) \right] .  \label{MOMCONS}
\end{equation}%
Using the constitutive relations in Eq. (\ref{CONST_REL}), 
 we recognize from Eq. (\ref{ECONS}) the energy flux $\mathbf{S}$  and the
energy density $U$  as 
\begin{equation}
\mathbf{S=}\frac{1}{4\pi }\mathbf{E\times B,\;\;\;\;\;\;\;\;}U\mathbf{=}%
\frac{1}{8\pi }(\mathbf{E}^{2}+\mathbf{B}^{2}),
\end{equation}%
while from Eq. (\ref{MOMCONS}) we obtain the momentum density $\mathbf{G}$
and  we identify the stress tensor $T_{ij}$ as, 
\begin{equation}
\mathbf{G}=\frac{1}{4\pi}\mathbf{E\times B,}\qquad T_{ij}=\frac{1}{8\pi }(%
\mathbf{E}^{2}+\mathbf{B}^{2})\delta _{ij}-\frac{1}{4\pi }%
(E_{i}E_{j}+B_{i}B_{j}).\mathbf{\;\;\;}
\end{equation}%
Outside the free sources, the conservation equations reads
\begin{equation}
\mathbf{\nabla \cdot S+}\frac{\partial U}{\partial t}=0,\;\;\;\;\;\;\;\;\;\;%
\frac{\partial G_{k}}{\partial t}+\partial _{i}T_{ik}=\frac{\alpha }{\pi }%
\left( E_{i}B_{i}\right) \partial _{k}\theta (z).
\end{equation}%
In other words, the stress tensor has the same form as in vacuum, but, as
expected, it is not conserved on the $\theta $-boundary because of the self-induced charge and current densities arising there.

Thus, the required expression for $T_{zz}\left( \Sigma ^{+}\right) \;$in Eq. (%
\ref{INTST}) is the standard one 
\begin{equation}
T_{zz}=\frac{1}{8\pi }\left[ E_{\parallel }^{2}-E_{z}^{2}+B_{\parallel
}^{2}-B_{z}^{2}\right] ,
\end{equation}%
where $E_{z}$ ($B_{z}$) denotes the electric (magnetic) field component
normal to the surface and $E_{\parallel }$ ($B_{\parallel }$) is the
component of the electric (magnetic) field parallel to the surface.
According to our results in the previous section, the electric and magnetic
fields for $z>0$ are
\begin{eqnarray}
\mathbf{E}\left( \mathbf{x}\right) &=&q\frac{\mathbf{x}-\mathbf{r}}{|\mathbf{%
x}-\mathbf{r}|^{3}}-q\frac{\tilde{\theta}^{2}}{4+\tilde{\theta}^{2}}\frac{%
\mathbf{x}+\mathbf{r}}{|\mathbf{x}+\mathbf{r}|^{3}},  \label{Efield} \\
\mathbf{B}\left( \mathbf{x}\right) &=&\frac{2q\tilde{\theta}}{4+\tilde{\theta%
}^{2}}\frac{\mathbf{x}+\mathbf{r}}{|\mathbf{x}+\mathbf{r}|^{3}}.
\label{Bfield}
\end{eqnarray}
Thus we find 
\begin{equation*}
\mathbf{F}=\frac{1}{4}\frac{q^{2}}{(4+\tilde{\theta}^{2})^{2}}\hat{\mathbf{e}%
}_{z}\int_{0}^{\infty }dR\frac{R}{\left( R^{2}+b^{2}\right) ^{3}}\left[
16R^{2}-(4+2\tilde{\theta}^{2})^{2}b^{2}+4\tilde{\theta}^{2}\left(
R^{2}-b^{2}\right) \right] =-\frac{q^{2}}{\left( 2b\right) ^{2}}\frac{\tilde{%
\theta}^{2}}{4+\tilde{\theta}^{2}}\hat{\mathbf{e}}_{z},
\end{equation*}%
in agreement with Eq. (\ref{force1}).

\subsection{Infinitely straight current-carrying wire near a planar $\theta$-boundary}

\label{infinite_wire}

Let us consider now an infinitely straight wire parallel to the $x$ axis
and carrying a current $I$ in the $+x$ direction. The wire is located in
vacuum ($\theta _{2} = 0$) at a distance $b$ from an infinite topological
medium with $\theta _{1} \neq 0$ in the region $z<0$. For simplicity we
choose the coordinates such that $y^{\prime }=0$. Therefore the current
density is 
$ j^\mu \left( \mathbf{x}^{\prime }\right) = I \eta ^{\mu} _{\; 1} \delta
\left( y^{\prime }\right) \delta \left( z^{\prime }-b\right)$.

The solution for this problem is given by 
\begin{equation}
A^{\mu }\left( \mathbf{x}\right) =I\int_{-\infty }^{+\infty }G_{\;1}^{\mu
}\left( \mathbf{x},\mathbf{r}\right) dx^{\prime } ,  \label{GenSolCurrent}
\end{equation}%
where $\textbf{x} - \mathbf{r}= \left( x - x^{\prime} \right) \hat{\mathbf{e}} _{x} + y 
\hat{\mathbf{e}} _{y} + \left( |z|+b\right) \hat{\mathbf{e}} _{z}$. Clearly
the nonzero component $A^{0}\left( \mathbf{x}\right) $ arising from the
GF implies that an electric field is induced. The required component of the GF, $G_{\;1}^{0}$, defined in Eq. (\ref{G0i}) is given by 
\begin{equation}
G_{\;1}^{0}\left( \mathbf{x},\mathbf{r}\right) =-\frac{2\tilde{\theta}}{4+%
\tilde{\theta}^{2}}\frac{y}{R^{2}}\left[ 1-\frac{|z|+b}{\sqrt{R^{2}+\left(
|z|+b\right) ^{2}}}\right] .  \label{01GreenCurrent}
\end{equation}%
Substituting Eq. (\ref{01GreenCurrent}) in Eq. (\ref{GenSolCurrent})
yields the electric potential,  which lacks an immediate interpretation. We can directly compute the electric field as $\mathbf{E}%
\left( \mathbf{x}\right) = - \nabla A ^{0} \left( \mathbf{x} \right)$, with
the result 
\begin{equation}
\mathbf{E}\left( \mathbf{x}\right) =\frac{4\tilde{\theta}I}{4+\tilde{\theta}%
^{2}}\left[ \frac{|z|+b}{y^{2}+\left( |z|+b\right) ^{2}}\mathbf{\hat{e}}_{y}-%
\frac{y\;\mbox{sign}\left( z\right) }{y^{2}+\left( |z|+b\right) ^{2}}\mathbf{%
\hat{e}}_{z}\right].
\end{equation}
We observe that the electric field for $z > 0$ is that due to a magnetic current located at $z = - b$, $\textbf{j} _{m,>} = - 4\tilde{\theta} I / (4+\tilde{\theta}^{2})  \hat{\textbf{e}} _{x}$. For $z < 0$ we have a magnetic current located at $z=b$ of the same strength $\textbf{j} _{m,<} = -  \textbf{j} _{m,>}$. Note that $\textbf{j} _{m,>}$ is antiparallel to the current of the wire, while $\textbf{j} _{m,<}$ is parallel.

Similarly we compute the magnetic field. This is 
\begin{equation}
\mathbf{B}\left( \mathbf{x}\right) =\nabla \times \left[ I\hat{\mathbf{e}}
_{i} \int_{-\infty }^{+\infty } G_{\;1}^{i} \left( \mathbf{x},\mathbf{r}%
\right) dx^{\prime }\right],
\end{equation}%
with $i=1,2$, where the corresponding GF are given by
Eqs. (\ref{Gij}). The result is 
\begin{equation}
\mathbf{B}\left( \mathbf{x}\right) = 2 I \mbox{sign}\left( z \right) \left[ -%
\frac{|b-z|}{y^{2}+\left( b-z\right) ^{2}}+\frac{\tilde{\theta}^{2}}{4+%
\tilde{\theta} ^{2}}\frac{\left( |z|+b\right) }{y^{2}+\left( |z|+b\right)
^{2}} \right] \mathbf{\hat{e}}_{y}+ 2Iy \left[ \frac{1}{y^{2}+\left(
b-z\right) ^{2}}- \frac{\tilde{\theta}^{2}}{4+\tilde{\theta}^{2}}\frac{1}{%
y^{2}+\left( |z|+b\right) ^{2}}\right] \mathbf{\hat{e}}_{z}.
\end{equation}
For $z > 0$ the magnetic field corresponds to the one produced by an image electric current  located at $z = - b$, flowing in the opposite direction to the current of the wire, $\textbf{j} _{e,>} = - 2\tilde{\theta} I / (4+\tilde{\theta}^{2})  \hat{\textbf{e}} _{x}$. For $z < 0$ we have an electric current located at $z=b$ of the same strength and flowing in the same direction of the current in the wire.

\section{Summary and outlook}

\label{summary}

Classical electrodynamics is a fascinating field theory on which a plethora
of technological devices rely. Advances in our theoretical understanding
ignite new technological developments and sometimes new discoveries demand
extending the limits of theories that lead to them. Chern-Simons forms and
topologically ordered materials are a good example of the above. In this
work we study a particular kind of Chern-Simons extension to   electrodynamics  that
consists of Maxwell Lagrangian supplemented by a parity-violating
Pontryagin invariant coupled to a scalar field $\theta$, restricted to the
case where $\theta$ is piecewise constant in different regions of space
separated by a common interface $\Sigma$.

It is well known that in this scenario the field equations in the bulk
remain the standard Maxwell equations but the discontinuity of $\theta$
alters the behavior of the fields at the interface $\Sigma$, giving rise to
effects such as: induced effective charge and currents at $\Sigma$ that are
determined by the fields at the interface, electric charges near a planar $%
\theta$-boundary induce magnetic mirror monopoles (and vice versa) and
nontrivial additional Faraday- and Kerr-like rotation of the plane of
polarization of electromagnetic waves traversing the interface $\Sigma$.

Here we focus on the Green's function method applied for the static case in $\theta$-electrodynamics. The method is illustrated by the case of a planar $\theta$-interface, 
where the corresponding Green's function  is calculated. The integral equation which defines the Green's function becomes  an algebraic equation due to the delta interaction arising in the $\theta$-boundary plus the symmetries present in the parallel directions. We show
how to compute the electromagnetic fields, on either side
of the interface from the Green's function.  Next we compute
the force between a point-like charge and a topological insulator. To this end we use the
Green's function to compute the interaction energy between  a
charge-current distribution and a $\theta$-boundary that mimics the topological insulator, with
vacuum energy removed. It can be shown that the above leads to the same
interaction force as that computed by momentum flux perpendicular to the
interface, for which the energy-momentum tensor and ensuing conservation
laws of $\theta$-electrodynamics were analyzed. Finally, we use the Green's
function  to obtain the electromagnetic fields  for an infinitely straight current-carrying wire parallel to the
interface.

For the case of the point-like charge in front of the $\theta$-interface, our 
results allow us to interpret the fields as those produced by the charge, its 
image, an induced magnetic monopole, and  a circulating current 
density at the interface, in agreement with previously existing results. 
Similarly the fields produced by the infinitely straight current-carrying wire 
and the $\theta$-boundary can be interpreted in terms of electric and 
magnetic current densities.

Let us emphasize that for or a given $\theta$-boundary, the fields produced by arbitrary external sources can be 
calculated once the Green's function is known. Our method can be applied to a broader kind of geometries determined
by the $\theta$-boundary. In fact, we can provide the Green's function for the
spherical and the cylindrical cases \cite{AMU_LARGO}. Given that our results depend on $\tilde{\theta} = \alpha (\theta _{1} - \theta _{2}) / \pi$, it is worth mentioning that they satisfy the quantum-mechanical periodicity condition $\theta \rightarrow \theta + 2 \pi n$, with $\theta=0, \pi$.

The Green's function method should also be useful for the extension to the 
dynamic case. In this respect, to our knowledge, little efforts have been done in 
the
context of topological insulators. Furthermore, Green's functions are also relevant for the computation of
other effects, such as the Casimir effect. Therefore, we expect our method
and results will be of considerable relevance and that they may constitute the basis for
numerous other researches.

\acknowledgments LFU acknowledges J. Zanelli for introducing him to the $\theta$-theories. LFU also thanks Alberto G\"{u}ijosa for useful comments and suggestions. M. Cambiaso  has been supported in part by the project FONDECYT (Chile) Initiation into Research Grant No. 11121633 and also wants to thank the kind hospitality at Instituto de Ciencias Nucleares, UNAM. LFU has been supported in part by the project No. IN104815
from Direcci\'on General Asuntos del Personal Acad\'emico (Universidad Nacional Aut\'onoma de M\'exico) and the project CONACyT (M\'exico) \# 237503. LFU and AMR thank the warm hospitality at Universidad Andres Bello.

\appendix

\section{GF for planar configuration in coordinate representation}

\label{IntegralsStaticCase}

\begin{figure}[tbp]
\begin{center}
\includegraphics[width = 2in]{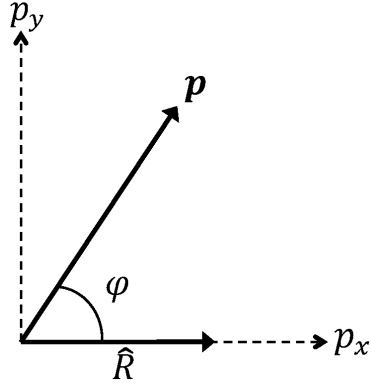}
\end{center}
\caption{{\protect\small $\mathbf{p}$-plane.}}
\label{planeint}
\end{figure}

Here we derive Eqs. (\ref{G00}-\ref{Gij}) by computing explicitly the
Fourier transform of the reduced GF, whose formula
we take from (\ref{GenSolGreenPlaneConf}). In the standard
case ($\tilde{\theta}=0$), the  reduced  vacuum  GF is \cite{CED} 
\begin{equation}
\mathfrak{g}\left( z,z^{\prime }\right) =\frac{1}{2p}e^{-p|z-z^{\prime }|}.
\label{gFree}
\end{equation}
In coordinate representation, the corresponding GF is obtained
by Fourier transforming (\ref{gFree}) as defined in Eq. (\ref{RedGreenDef}), 
\begin{equation}
\mathcal{G}\left( \mathbf{x},\mathbf{x}^{\prime }\right) =4\pi \int \frac{%
d^{2}\mathbf{p}}{\left( 2\pi \right) ^{2}}e^{i\mathbf{p}\cdot \left( \mathbf{%
x}-\mathbf{x}^{\prime }\right) _{\parallel }}\frac{1}{2p}e^{-p|z-z^{\prime }|}.
\label{FTgFree}
\end{equation}%
This double integral becomes easier to perform if we express the area
element in polar coordinates, $d^{2}\mathbf{p}%
=pdpd\varphi $ (instead of the Cartesian ones), and choose the $p_{x}$-axis
in the direction of the vector $\mathbf{R}=\left( \mathbf{x}-\mathbf{x}%
^{\prime }\right) _{\parallel }$, as shown in Fig. \ref{planeint}.
Noting that $\mathbf{p}\cdot \mathbf{R}=pR\cos \varphi $, we can write 
\begin{equation}
\mathcal{G}\left( \mathbf{x},\mathbf{x}^{\prime }\right) =\int_{0}^{\infty
}dpe^{-p|z-z^{\prime }|}\left\{ \frac{1}{2\pi }\int_{0}^{2\pi }e^{ipR\cos
\varphi }d\varphi \right\} .  \label{FTgFree2}
\end{equation}%
where $R=|\left( \mathbf{x}-\mathbf{x}^{\prime }\right) _{\parallel }|$. The
braces in this equation enclose an integral representation of the Bessel
function $J_{0}\left( pR\right) $. The resulting integral, 
\begin{equation}
\mathcal{G}\left( \mathbf{x},\mathbf{x}^{\prime }\right) =\int_{0}^{\infty
}J_{0}\left( pR\right) e^{-p|z-z^{\prime }|}dp,  \label{FTgFree3}
\end{equation}%
is well-known, see for example Ref. \cite{Gradshteyn}. The final result is 
\begin{equation}
\mathcal{G}\left( \mathbf{x},\mathbf{x}^{\prime }\right) =\frac{1}{\sqrt{%
R^{2}+|z-z^{\prime }|^{2}}}=\frac{1}{|\mathbf{x}-\mathbf{x}^{\prime }|},
\label{FTgFree4}
\end{equation}%
which is the  vacuum  GF in coordinate representation \cite{CED}.
In the following we use a similar procedure to compute the required integrals
for establishing Eqs. (\ref{G00}-\ref{G0i}). We first consider the component$\;G_{\;0}^{0}$. From Eq. (\ref%
{GenSolGreenPlaneConf}) we find 
\begin{equation}
g_{\;0}^{0}\left( z,z^{\prime }\right) =\mathfrak{g}\left( z,z^{\prime
}\right) +A\left( z,z^{\prime }\right) p ^{2}\tilde{\theta}\mathfrak{g}\left(
a,a\right) ,  \label{staticg00}
\end{equation}%
where the function $A\left( z,z^{\prime }\right) $ is 
\begin{equation}
A\left( z,z^{\prime }\right) =-\frac{\tilde{\theta}}{4+\tilde{\theta}^{2}}%
p ^{-2}e^{-pZ},  \label{A-static}
\end{equation}%
with the notation $Z=|z-a|+|z^{\prime }-a|$. In this way, the component $%
G_{\;0}^{0}$ is given by 
\begin{equation}
G_{\;0}^{0}\left( \mathbf{x},\mathbf{x}^{\prime }\right) =\mathcal{G}\left( 
\mathbf{x},\mathbf{x}^{\prime }\right) -\frac{\tilde{\theta}^{2}}{4+\tilde{%
\theta}^{2}}\int_{0}^{\infty }J_{0}\left( pR\right) e^{-pZ}dp,
\label{Fourier00}
\end{equation}%
in coordinate representation. As before we use the integral representation
of the Bessel function $J_{0}\left( pR\right) $ to perform the angular
integration. The resulting integral is the same as in (\ref{FTgFree3}), thus
we obtain 
\begin{equation}
G_{\;0}^{0}\left( \mathbf{x},\mathbf{x}^{\prime }\right) =\frac{1}{|\mathbf{x%
}-\mathbf{x}^{\prime }|}-\frac{\tilde{\theta}^{2}}{4+\tilde{\theta}^{2}}%
\frac{1}{\sqrt{R^{2}+Z^{2}}}. \label{G00Coordinates}
\end{equation}

Now we evaluate the components $G_{\;1}^{0}$ and $G_{\;2}^{0}$. The corresponding reduced GF are
\begin{equation}
g_{\; i}^{0}\left( z,z^{\prime }\right) = - i \epsilon _{0ij3} p ^{j} A\left( z,z^{\prime }\right) ,  \label{staticg01-g02}
\end{equation}
with $A\left( z,z^{\prime }\right) $ given by (\ref{A-static}). For
 convenience  we define the vector 
\begin{equation}
\mathbf{I} \left( \mathbf{x},\mathbf{x}^{\prime }\right) =(I ^{1} , I ^{2}) = 4\pi \int \frac{%
d^{2}\mathbf{p}}{\left( 2\pi \right) ^{2}}e^{i\mathbf{p}\cdot \left( \mathbf{%
x}-\mathbf{x}^{\prime }\right) _{\parallel }}\mathbf{p\;}%
p^{-2}e^{-pZ},  \label{vectorI}
\end{equation}%
with $\mathbf{p}=\left( p_{x},p_{y}\right) $, in terms of which we have%
\begin{equation}
G_{\; i}^{0}\left( \mathbf{x},\mathbf{x}^{\prime }\right) =i \frac{\tilde{\theta}}{4+\tilde{\theta}^{2}} \epsilon _{0ij3} I ^{j} \left( \mathbf{x},\mathbf{x}^{\prime }\right) .
\end{equation}
We calculate the integral (\ref{vectorI}) in the same coordinate system as
before (see Fig. \ref{planeint}), and then we rewrite the result in a
vector form. The integral can be written as 
\begin{equation}
\mathbf{I}_{p}\left( \mathbf{x},\mathbf{x}^{\prime }\right) = 2 \int_{0}^{\infty }dpe^{-pZ}\left\{ \frac{1}{2\pi }\int_{0}^{2\pi }\left[ 
\begin{array}{c}
\cos \varphi \\ 
\sin \varphi%
\end{array}%
\right] e^{ipR\cos \varphi }d\varphi \right\} ,  \label{vectorI2}
\end{equation}%
where the subscript $p$ indicates that the vector$\;\mathbf{p}$ is written
in the particular coordinate system of Fig. \ref{planeint}. Both the
required angular and radial integrals are well-known and the result is 
\begin{equation}
\mathbf{I}_{p}\left( \mathbf{x},\mathbf{x}^{\prime }\right) = 2i \widehat{\mathbf{R}}\int_{0}^{\infty }J_{1}\left( pR\right) e^{-pZ}dp=\frac{2 i}{R}\left( 1-\frac{Z}{\sqrt{R^{2}+Z^{2}}}\right) \widehat{\mathbf{R}}.
\label{vectorI3}
\end{equation}%
As a consequence of the chosen coordinate system we find that  $I_2=0$, in such a way that  the vector $%
\mathbf{I}_{p}$ becomes parallel to $\widehat{\mathbf{R}}$. However this can
be generalized in a direct way to an arbitrary coordinate system as 
\begin{equation}
\mathbf{I}\left( \mathbf{x},\mathbf{x}^{\prime }\right) = 2i \frac{\textbf{R}}{%
R^{2}}\left( 1-\frac{Z}{\sqrt{R^{2}+Z^{2}}}\right) .
\end{equation}%
Thus we find 
\begin{eqnarray}
G_{\; i}^{0}\left( \mathbf{x},\mathbf{x}^{\prime }\right) = - \frac{2 \tilde{\theta}}{4+\tilde{\theta}^{2}} \frac{\epsilon _{0ij3} R ^{j}}{R ^{2}}\left( 1-\frac{Z}{\sqrt{R^{2}+Z^{2}}}\right) .
\end{eqnarray}
In order to evaluate the components $G ^{i} _{\; j}$ we first observe that the corresponding reduced GF can be written as
\begin{equation}
g ^{i} _{\; j} \left( z , z ^{\prime} \right) = \eta ^{i} _{\; j} g ^{0} _{\; 0} \left( z , z ^{\prime} \right) + \tilde{\theta} \mathfrak{g} (a,a) A \left( z , z ^{\prime} \right) p ^{i} p _{j} , \label{staticgij}
\end{equation}
where $g ^{0} _{\; 0}$ is given by Eq. (\ref{staticg00}). Now we need to compute the Fourier transformation of Eq. (\ref{staticgij}) as defined in Eq. (\ref{RedGreenDef}).  However the first term was studied before and the result is given by Eq. (\ref{G00Coordinates}), thus leading to study only the last term. To this end we introduce the vector
\begin{equation}
\mathbf{K}\left( \mathbf{x},\mathbf{x}^{\prime }\right) = ( K ^{1} , K ^{2} ) = 4 \pi \int \frac{d^{2}\mathbf{p}}{\left( 2\pi \right) ^{2}}e^{i\mathbf{p}\cdot \left( \mathbf{x}-\mathbf{x}^{\prime }\right) _{\parallel }}\frac{\mathbf{p}}{p} p^{-2}e^{-pZ},  \label{Kvector1}
\end{equation}
from which the required integral will be calculated by taking the spatial derivative.  The integral (\ref{Kvector1})\ can be computed again in the
particular coordinate system of the Fig. \ref{planeint}. In the polar
coordinates defined in the $\mathbf{p}$-plane the integral reads 
\begin{equation}
\mathbf{K}_{p}\left( \mathbf{x},\mathbf{x}^{\prime }\right)
=2\int_{0}^{\infty }\frac{dp}{p}e^{-pZ}\left\{ \frac{1}{2\pi }\int_{0}^{2\pi
}\left[ 
\begin{array}{c}
\cos \varphi \\ 
\sin \varphi%
\end{array}%
\right] e^{ipR\cos \varphi }d\varphi \right\} . \label{Kaux}
\end{equation}
Note that the braces in this equation enclose an integral representation of
the Bessel function $J_{1}\left( pR\right) $. The resulting integral is well-known and the final result is 
\begin{equation}
\mathbf{K}_{p}\left( \mathbf{x},\mathbf{x}^{\prime }\right)
=2 i \int_{0}^{\infty }\frac{dp}{p}J_{1}\left( pR\right) e^{-pZ}\hat{\mathbf{R}} = 2i\frac{\sqrt{%
R^{2}+Z^{2}}-Z}{R}\hat{\mathbf{R}},  \label{Kvector4}
\end{equation}
where $\hat{\mathbf{R}}$ is the unit vector shown in Fig. \ref{planeint}.
The generalization to an arbitrary coordinate system is then 
\begin{equation}
\mathbf{K}\left( \mathbf{x},\mathbf{x}^{\prime }\right) =2i\frac{\sqrt{%
R^{2}+Z^{2}}-Z}{R^{2}} \textbf{R} .  \label{Kvector5}
\end{equation}
Note that the required integral involve the term $p ^{i} p _{j}$ which can be generated from (\ref{Kvector1}) as follows 
\begin{eqnarray}
i \partial _{j} K ^{i } \left( \mathbf{x},\mathbf{x}^{\prime }\right) = 4 \pi \int \frac{d^{2} \mathbf{p}}{\left( 2\pi \right) ^{2}}e^{i\mathbf{p}\cdot \left( \mathbf{x}-\mathbf{x}^{\prime }\right) _{\parallel }}\frac{p ^{i} p _{j}}{p} p^{-2}e^{-pZ} .
\end{eqnarray}%
By using the final form of $\mathbf{K}\left( \mathbf{x},\mathbf{x}^{\prime
}\right) $, given by Eq. (\ref{Kvector5}), one can further check the
consistency condition $ \partial _{1} K ^{2}\left( \mathbf{x},\mathbf{x} ^{\prime }\right) = \partial _{2} K ^{1}\left( \mathbf{x},\mathbf{x} ^{\prime }\right)$ required by the cross terms involving $p ^{1} p _{2} = p ^{2} p _{1} = - p _{x} p _{y}$. From the previous results, the $G ^{i} _{\; j}$ components of the GF matrix in coordinate representation can be written
as 
\begin{eqnarray}
G _{\; j}^{ i }\left( \mathbf{x},\mathbf{x}^{\prime }\right) &=& \eta ^{i} _{\; j} G _{\; 0}^{ 0}\left( \mathbf{x},\mathbf{x}^{\prime }\right) - \frac{i}{2}\frac{\tilde{\theta}^{2}}{4+ \tilde{\theta}^{2}} \partial _{j} K ^{i} \left( \mathbf{x},\mathbf{x}^{\prime }\right) .
\end{eqnarray}
These results establish Eqs. (\ref{Gij}).

\end{document}